\documentclass[11pt,twoside]{article}
\usepackage{asp2014}

\aspSuppressVolSlug

\resetcounters

\begin{document}

\title{A Quantitative Comparison of Opacities Calculated Using the Distorted-Wave and $\boldsymbol{R}$-Matrix Methods}

\author{F.~Delahaye,$^1$ N.~R.~Badnell,$^2$ C.~P.~Ballance,$^3$ P.~Palmeri,$^4$ S.~Preval,$^2$
P.~Quinet,$^4$ C.~Ramsbottom,$^3$ R.~T.~Smyth,$^3$ M.~Turkington,$^3$ and C.~J.~Zeippen$^1$\\
\affil{$^1$LERMA, Observatoire de Paris, PSL Research University, CNRS, Sorbonne University, UPMC Univ Paris 06, F-92190 Meudon, France; \email{franck.delahaye@obspm.fr}}
\affil{$^2$Department of Physics, University of Strathclyde, Glasgow G4 0NG, UK}
\affil{$^3$CTAMOP, School of Mathematics and Physics, Queen's University, Belfast BT7 1NN, UK}
\affil{$^4$Physique Atomique et Astrophysique, Universit\'e de Mons - UMONS, B-7000 Mons, Belgium}
}


\paperauthor{F.~Delahaye}{franck.delahaye@obspm.fr}{}{Observatoire de Paris}{LERMA}{Meudon}{}{F-92190}{France}
\paperauthor{C.~J. Zeippen}{}{}{Observatoire de Paris}{LERMA}{Meudon}{}{F-92190}{France}
\paperauthor{N.~R.~Badnell}{}{}{University of Strathclyde}{Department of Physics}{Glasgow}{}{G4 0NG}{UK}
\paperauthor{S.~Preval}{}{}{University of Strathclyde}{Department of Physics}{Glasgow}{}{G4 0NG}{UK}
\paperauthor{C.~P.~Ballance}{}{}{Queen's University}{CTAMOP, School of Mathematics and Physics}{Belfast}{}{BT7 1NN}{UK}
\paperauthor{C.~Ramsbottom}{}{}{Queen's University}{CTAMOP, School of Mathematics and Physics}{Belfast}{}{BT7 1NN}{UK}
\paperauthor{R.~T.~Smyth}{}{}{Queen's University}{CTAMOP, School of Mathematics and Physics}{Belfast}{}{BT7 1NN}{UK}
\paperauthor{M.~Turkington}{}{}{Queen's University}{CTAMOP, School of Mathematics and Physics}{Belfast}{}{BT7 1NN}{UK}
\paperauthor{P.~Palmeri}{}{}{Universit\'e de Mons - UMONS}{Physique Atomique et Astrophysique}{Mons}{}{B-7000}{Belgium}
\paperauthor{P.~Quinet}{}{}{Universit\'e de Mons - UMONS}{Physique Atomique et Astrophysique}{Mons}{}{B-7000}{Belgium}


\begin{abstract}
  The present debate on the reliability of astrophysical opacities has reached a new climax with the recent measurements of Fe opacities on the Z-machine at the Sandia National Laboratory  \citep{Bailey2015}. To understand the differences between theoretical results, on the one hand, and experiments on the other, as well as the differences among the various theoretical results, detailed comparisons are needed. Many ingredients are involved in the calculation of opacities; deconstructing the whole process and comparing the differences at each step are necessary to quantify their importance and impact on the final results. We present here such a comparison using the two main approaches to calculate the required atomic data, the $R$-Matrix and distorted-wave methods, as well as sets of configurations and coupling schemes to quantify the effects on the opacities for the \ion{Fe}{xvii} and \ion{Ni}{xiv} ions.
\end{abstract}

\section{Introduction}
\label{Introduction}

More than three decades ago \citet{Simon1982} made a plea for a recalculation of metal opacities to solve the so-called
{\it Cepheids problem}. Two groups answered this call: the team led by Forrest Rogers and Carlos Iglesias from the Lawrence Livermore National Laboratory (LLNL), referred to as OPAL, who used an equation of state (EOS) based on a ``physical picture'' \citep{Rogers1996}, and the international Opacity Project (OP) led by Mike Seaton who preferred an EOS approach based on a ``chemical picture'' \citep{ADOC_II}. A succinct summary of these earlier days can be found in a review by \citet{Zeippen1995}.

By the mid 1990s both groups ended up with enhanced metal opacities: the factor of 2 to 3 expected by \citeauthor{Simon1982} was indeed obtained. The OPAL physical approach enabled careful inclusion of plasma effects and a detailed EOS \citep{OPAL1992}. The chemical approach favored detailed calculations of the basic atomic data using state-of-the-art codes such as {\sc superstructure} \citep{Eissner1974, Nussbaumer1978}, CIV3 \citep{hib75}, and the $R$-Matrix suite of scattering codes \citep{Burke1971, ADOC_I,  ADOC_II, Hummer1993}, the plasma effects were then added as perturbations \citep{ADOC_I, ADOC_II}. Since then, these opacity datasets have been successfully incorporated in stellar models replacing the previous standard, namely, the Astrophysical Opacity Library \citep{Cox1960, Cox1965, Cox1970a, Cox1970b, Cox1976}.

Although the two independent projects were based on largely different physical frameworks, the agreement between the OPAL and OP opacities after the latter were upgraded \citep[][OP2005 hereafter]{Badnell2005} was satisfactory in solar interior conditions, and their public datasets rapidly became reference data in stellar astrophysics. The format of the OPAL Rosseland-mean opacity tables\footnote{\url{https://opalopacity.llnl.gov/}} became the de facto standard, and the OP basic atomic data and monochromatic opacities have been useful in many astrophysical applications (line identifications, synthetic spectra synthesis, etc.). In more recent years, the OP has made available two online servers\footnote{\url{http://cdsweb.u-strasbg.fr/topbase/home.html}} with opacity and radiative acceleration data and utilities \citep{Mendoza2007, Delahaye2016}.

Despite the good agreement between the OPAL and OP opacities, as well as with other more recent efforts such as OPAS from the CEA \citep{Blancard2012,Mondet2015} and LEDCOP (now ATOMIC) from Los Alamos National Laboratory \citep{Colgan2013b, Colgan2013a, Colgan2015, Colgan2016} and the accurate reproduction of the helioseismic benchmarks by standard solar models using them, the integrity of the opacity tables was questioned in 2004 when \citet{Asplund2004,Asplund2005} recommended revised solar abundances with reduced C, N, and O contents at the level of 30--40\%. The new solar composition seriously deteriorated the much coveted agreement between theory and helioseismic measurements.

In the last decade a large number of studies have appeared suggesting further opacity changes \citep[for recent developments, see][and references therein]{Villante2014,Serenelli2017}; such changes are viewed by many as a way to reconcile the new solar composition with the helioseismic benchmarks. There is indeed a direct link between opacities and composition \citep[see, for example,][]{TurckChieze1988}; however, the problem is still unresolved as opacity producers cannot find a well-defined cause or source for such increases. Furthermore, the revised solar abundances are still debated, and independent groups who have developed other 3D stellar atmosphere modelling codes do not entirely agree with the predictions by \citet{Asplund2005}. The latter have themselves reassessed their initial predictions increasing somewhat the abundance of C, N and O as a result \citep{Asplund2009}. There is still room for improvement and the low-$Z$ composition is still under debate. In a few words, the central question is, opacities or composition?

Later on experimentalists have become involved in the opacity debate, and recent measurements on the Z facility at Sandia National Laboratories by \citet{Bailey2015} certainly suggest a revision of the theoretical opacities. However, theorists are still puzzled as to what is really missing in their calculations or which approximations need to be relaxed. Whichever the approach (the physical or chemical picture), EOS, or atomic configuration expansions, most theoretical efforts end up short of the experiment. This situation has led us to explore the fine details of opacity calculations comparing various approaches. Recent debates among producers \citep{Nahar2016a, Nahar2016b, Blancard2016, Iglesias2017} have emphasized the necessary compromise between accuracy and completeness.

In the present work we explore the differences between the OP coupled-channel method and the perturbative methods used by most other approaches to calculate the basic atomic data (energy levels, $f$-values,  photoionization cross sections) required to derive opacities. More specifically, while the latter is simpler and allows for extensive sets of electron configurations,  the former treats resonances in a formal {\it ab initio} manner although the implementation of large sets of configurations rapidly becomes computationally intractable. A quantitative evaluation is required. We have been evaluating the effects of the different resonance treatments on the accuracy of photoionization cross sections using the atomic structure code {\sc autostructure}  \citep[AS,][]{Badnell2011} for the larger sets of configurations and the $R$-Matrix suite  for accurate resonance representations. We present here preliminary results for \ion{Fe}{xvii} and \ion{Ni}{xiv}.

\section{Opacities: Ingredients}
\label{Opacities}

Many ingredients enter in the calculation of opacities: (i) the state of the absorbing material, that is, the ionic fraction of the constituents and the level populations for each ion that we refer to as the EOS; (ii) the absorbing coefficients associated with the different processes (photoexcitation, scattering, and photoionization), i.e., the raw atomic data; (iii) the ion--ion and electron--ion interactions; and (iv) the physical conditions of the plasma whereby atoms cannot be considered isolated, and are treated via line broadening.

Many differences exist among current theoretical approaches at all levels; nonetheless, it appears that the Rosseland mean opacities for the solar mixture agree to within $4\%$ between OPAL, OPAS, OP, and LEDCOP \citep[see, for example, OP2005,][]{Delahaye2006, Blancard2012}. Naturally some compensation is introduced in the process of mixing different elements and averaging over frequencies, and the finer details in the monochromatic opacities for the different elements do not show the same level of agreement \citep{Blancard2012}.

When we look more closely at the contributions from the different processes of each element at each frequency point, the combination of the different processes gives the total elemental contribution
\begin{equation}
\label{Eq1}
\kappa_{\rm tot}(\nu) = \kappa_{\rm bound-bound}(\nu) + \kappa_{\rm bound-free}(\nu) + \kappa_{\rm free-free}(\nu)\,.
\end{equation}
To derive the Rosseland mean we have to sum up these contributions; depending on the element and the physical conditions, the contribution of each of these processes varies. In stellar interiors and especially in the Sun, the photoionization contribution is dominant for all metals as shown, for instance, by \citet{Blancard2012}. In our current endeavor to quantify all uncertainties in opacity calculations, it is then a good starting point to examine the photoionization cross sections by different methods and their impact on the derived opacities.

For the conditions we are interested in, akin to the boundary of the solar convection zone and the Sandia experiment ($T=2.1\times10^6$~K and $\rho=3\times10^{22}$~cm$^{-3}$), there are two main pathways to photoionize an atom: the direct process whereby a photon is absorbed ejecting a free electron, and an indirect mechanism where the photon excites the electron to an autoionizing state (also called a {\it quasi-bound state}), which then autoionizes faster than by radiative decay (see Section~\ref{DW}). A key difference between perturbative and coupled-channel methods lies in the treatment of this indirect channel. While the former methods treat resonances as individual lines, the OP $R$-Matrix formalism takes into account the interference between the direct and indirect channels giving rise to Beutler--Fano resonance profiles \citep{Fano1961}.

In the present work we choose to study this interference and disentangle its effects rather than the impact of the convergence of the configuration expansions.

\subsection{$\boldsymbol{R}$-Matrix Method}
\label{$R$-matrix}

Photoionization cross sections have been calculated using the Breit--Pauli $R$-Matrix method \citep{Burke1971, ADOC_I,  ADOC_II, Hummer1993} as used in the Iron Project (IP) and in a number of previous publications. The aims and methods of the IP are presented in  \citet{Hummer1993}; we briefly summarize its main features as follows. In the close-coupling (CC) approximation the wave-function expansion $\Psi(E)$ for a total angular symmetry $J \pi$ of the $(N + 1)$-electron system is represented in terms of the target ion states as
\begin{equation}
\label{Eq2}
\Psi (E) = A \sum_{i} \chi _i \Theta _i + \sum_{j} c_j \Phi _j\,,
\end{equation}
where $\chi _i$ is the target ion wave function in a specific level $J_i\pi _i$. $\Theta _i$ is the wave function for the $(N + 1)$th electron in a channel labelled as
\begin{equation}
\label{Eq3}
S_i L_i(J_i) k_i ^2 l_i \pi _i(J\pi)
\end{equation}
with $k_i ^2$ being its incident kinetic energy. In the second sum $\Phi _j$ are correlation wave functions of the $(N + 1)$-electron system that (i) compensate for the orthogonality conditions between the continuum and the bound orbitals, and (ii) represent additional short-range correlations that are often of crucial importance in scattering and radiative CC calculations for each symmetry. The $\Phi _j$ are also referred to as the bound channels, as opposed to the continuum or free channels
in the first summation over the target states. Details of the diagonalization of the Breit-Pauli $R$-Matrix method are given in many previous reports \citep[e.g.,][]{Berrington1995}.

\subsection{IPIRDW Method}
\label{DW}

The photoionization cross sections of the bound states of Ni and Fe can be generated using the AS atomic structure code  \citep{Badnell2011}. AS computes term energies, fine-structure level energies, radiative and Auger rates,
photoionization cross sections, and electron-impact collision strengths. In these calculations the atomic wave functions are constructed by diagonalizing the non-relativistic Hamiltonian whereby the radial orbitals are determined using a scaled statistical model potential. $LS$ terms are represented by configuration-interaction (CI) wave functions of the type
\begin{equation}
\label{Eq4}
\Psi(LS)= \sum_i~c_i~\Phi_i\,.
\end{equation}
Continuum wave functions are constructed within the distorted-wave (DW) approximation. Relativistic fine-structure levels and rates can be obtained by diagonalizing the Breit--Pauli Hamiltonian in IC. The one- and two-body operators have been fully implemented to order $\alpha^2 Z^4$, where $\alpha$ is the fine-structure constant and $Z$ the atomic number.

The non-resonant direct photoionization process for an element {\it X}
\begin{equation}
\label{Eq5}
    X^{n+} + h\nu \rightarrow X^{(n+1)+} + e^-
\end{equation}
is treated separately from the photoexcitation--autoionization process
\begin{equation}
\label{Eq6}
    X^{n+} + h\nu \rightarrow  (X^{n+})^* \rightarrow X^{(n+1)+} + e^-
\end{equation}
within the framework of the IPIRDW approximation \citep{BadnellSeaton2003}.

\section{Results}

In detailed quantitative comparisons between datasets, it is important to clearly establish both the magnitude and origin of the discrepancies that are found. In the case of opacities, the origin, i.e., the atomic data, EOS, or broadening,
must be ascertained with great care. For example, the differences in the atomic data can be multiple: the method (DW {\it vs}. $R$-Matrix), the specific set of configurations, or the coupling scheme.

\subsection{OP Opacity Sets}
\label{opsets}

The first set of OP opacities (hereafter OP1995) was delivered in the mid 1990s \citep{Seaton1994, Seaton1995, Zeippen1995}, and included atomic data calculated using the $R$-Matrix suite of codes in $LS$ coupling complemented with $f$-values computed with the atomic structure code {\sc superstructure} for selected Fe ions \citep{LinasGray1995}. A final and more complete dataset containing additional inner-shell contributions calculated in the DW approximation with AS was released by OP2005.

\citet[][hereafter NP2016]{Nahar2016a} recently presented new $R$-Matrix results on the photoionization of \ion{Fe}{xvii}. In their Fig.~3 they highlight the enhancements in the photoionization cross sections resulting from 30-state and 99-state  $LS$ calculations when compared to the 2-state $LS$ cross section from OP1995. It must be noted that the 2-state OP data only include valence--electron absorption, while in the second release by OP2005, contributions from inner-shell photoexcitation and photoionization were taken into account in the DW approximation neglecting the interference between the direct and indirect channels mentioned in Section~\ref{DW}.

We show in Fig.~\ref{fig1} that the background cross sections by OP2005 and NP2016 for the ${\rm 2p^53d\ ^1D^o}$ and ${\rm 2p^53p\ ^1S^e}$ states of \ion{Fe}{xvii} are in close agreement. The large enhancement therefore is more a consequence of inner-shell photoionization rather than channel interference. On the other hand, the detailed resonance profiles are different as they are assumed symmetric in DW in contrast to the more realistic Fano profile associated with channel interference in the $R$-Matrix method. This will have an impact on the final opacities as shown by \citet{Delahaye2013}: a 10\% increase was found in $\kappa_{R}$ from the photoionization contribution of \ion{Ni}{xiv} when using the $R$-Matrix approach, part of which is due to CI between complexes and part from the $R$-Matrix approach.

\articlefiguretwo{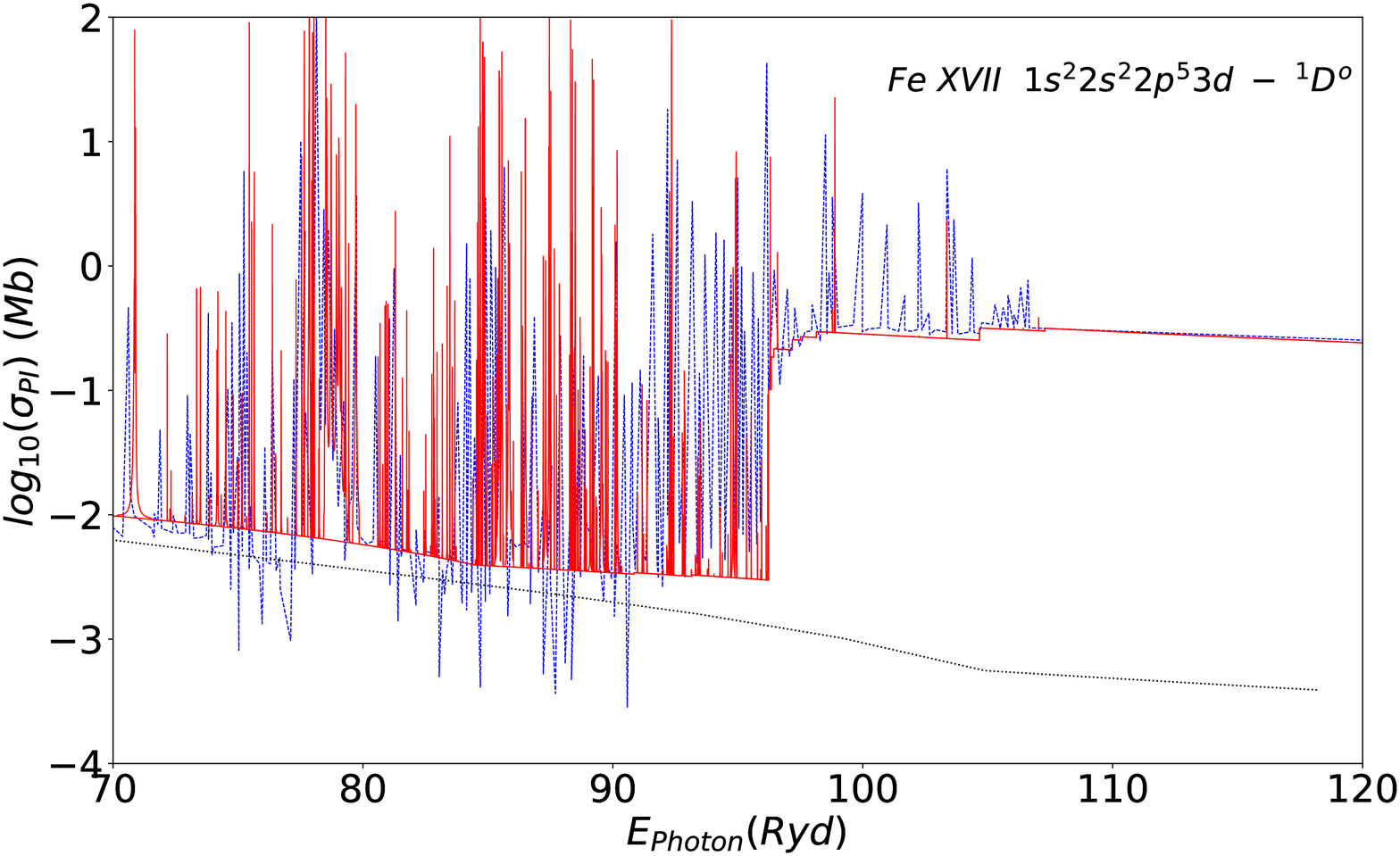}{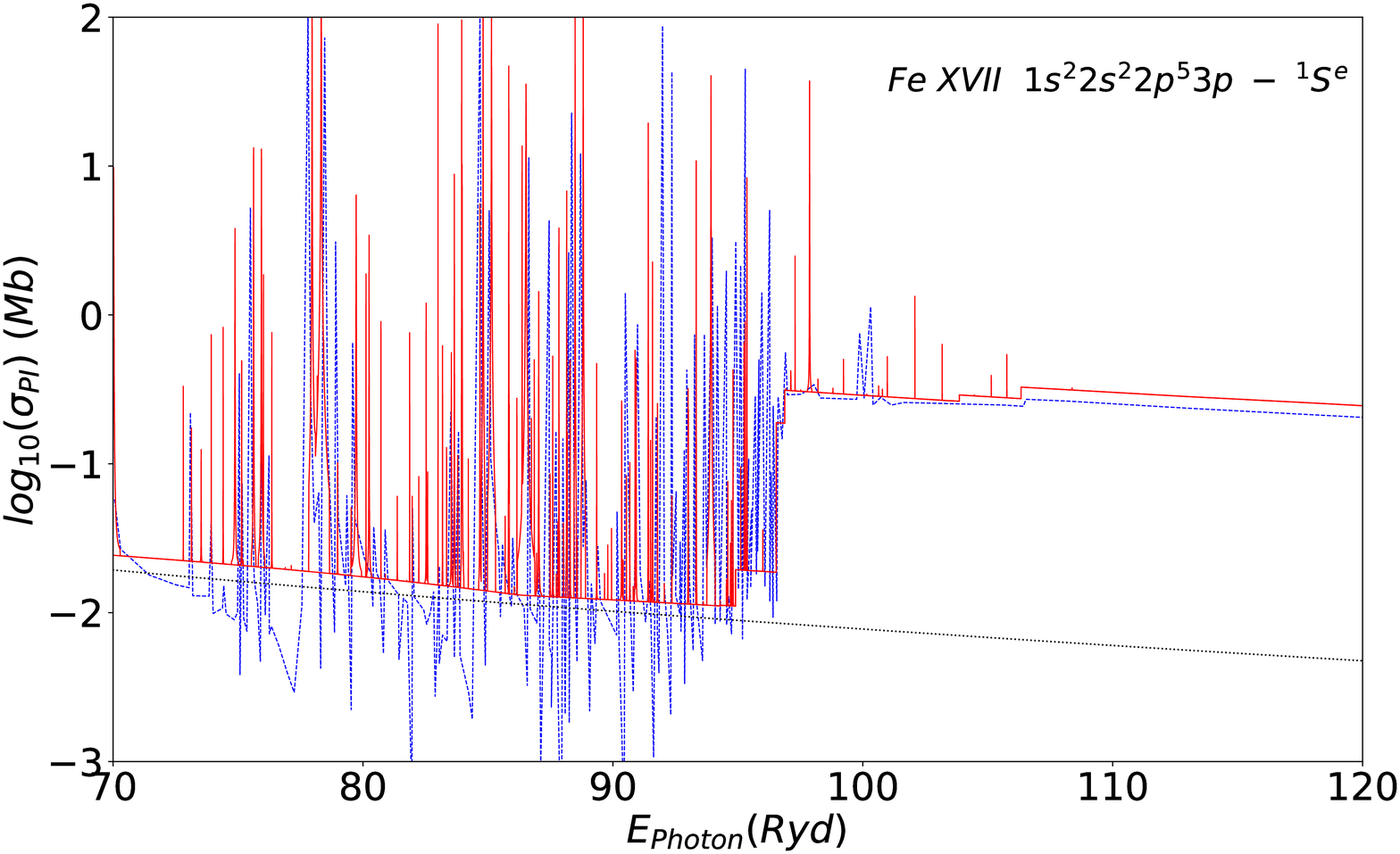}{fig1}{Photoionization cross section $\sigma_{\rm PI}$ of the  ${\rm 1s^22s^22p^53d\ ^1D^o}$ ({\it left panel}) and ${\rm 1s^22s^22p^53p\ ^1S^e}$ ({\it right panel}) states of \ion{Fe}{xvii}. Blue curve: \citet{Nahar2016a}. Black curve: 2-state (TOPbase). Red: OP2005.}

\subsection{New OP Opacity Codes}

In the OP approach the different steps to calculate opacities involve: (i) obtaining the raw atomic data (levels, photoionization cross sections, $f$-values); (ii) determining the ionic fractions and level populations with the EOS; and (iii) applying the different broadening mechanisms (natural, electron impact broadening) to the bound--bound transitions and bound--free resonances. In order to clearly establish the impact of each of these steps on the final opacity we have developed a new set of codes allowing us to use different atomic data sources. We can then download data directly from databases such as TOPbase\footnote{\url{http://cdsweb.u-strasbg.fr/topbase/topbase.html}}  \citep{Cunto1993}, NORAD\footnote{\url{http://norad.astronomy.ohio-state.edu/}} or, if complete and consistent, from any atomic database linked to VAMDC \citep{VAMDC2014, VAMDC2016}, as well as direct output from AS or the $R$-Matrix suite of codes. It is
then easier to establish clearly the impact due to the  method (DW {\it vs.} close-coupling), target representation (configuration-interaction expansion), coupling ($LS$ {\it vs.} IC), etc. Our new codes implement a finer frequency mesh for generating monochromatic opacities that is crucial to determine radiative accelerations. In the present work all monochromatic opacities have been calculated using a mesh with $10^5$ frequency points.

Our starting point is the comparison between the Rosseland means from different theoretical groups shown in Fig.~3 of \citet{Bailey2015} and summarized in Table~1 \citep[based on Table~1 from][]{Blancard2016}. It is difficult to disentangle the possible discrepancy sources, even if some new contributions (new broadening, extended set of configurations) are
invoked in, for example, \citet{Blancard2016}, \citet{Nahar2016a}, and \citet{Iglesias2017}.

\begin{table}
  \caption{Comparison of the \ion{Fe}{xvii} Rosseland mean opacity (in cm$^2$\,g$^{-1}$) from different calculations \citep[based on numbers from][]{Blancard2016}. The OP2005 release is taken as reference \citep[for details, see][]{Blancard2016}.}\label{tab1}
  \smallskip
  {\centering \small
  \begin{tabular}{lll}
  \tableline
  \noalign{\smallskip}
  Method & $\kappa_R\times 0.196$ & $\kappa_R / $ OP2005 \\
  \noalign{\smallskip}
  \tableline
  \noalign{\smallskip}
  OP2005  & 126.06 &1.00\\
  NP2016  & 170.18 &1.35\\
  ATOMIC  & 166.40 &1.32\\
  OPAS    & 195.39 &1.55\\
  SCO-RCG & 172.70 &1.37\\
  SCRAM   & 160.10 &1.27\\
  TOPAZ   & 152.53 &1.21\\
  \noalign{\smallskip}
  \tableline
  \end{tabular}

  }
\end{table}

\subsection{Coupling Scheme, Resolution, Extended Targets, and Inner Shells}

All the following results have been obtained at the plasma conditions of Sandia experiment ($T=2.1\times10^6$\,K and $\rho=3\times 10^{22}$\,cm$^{-3}$). We test each effect independently and start by reanalysing the data used in OP2005. To test the impact of the frequency mesh resolution, we increase the mesh step from $10^4$ to $10^5$ points finding an increment in the Rosseland mean $\kappa_R$ of at the most $1\%$. This outcome concurs with \citet{Seaton1994}.

We find a 2\% reduction in $\kappa_R$ when switching from $LS$ coupling to IC. Although the effect of this switch was investigated by \citet{BadnellSeaton2003} for the 6-element solar mixture, the final updated OP2005 data used $LS$ coupling throughout. It should be mentioned that OP2005 did not include the required renormalization of the bound-state populations taking into account the population redistribution to the autoionizing levels that occurs at high density; this leads to a 2\% overestimate of $\kappa_R$ for \ion{Fe}{xvii} under the present conditions. Moreover, the coupling scheme and the renormalization in the present results are certainly linked, as the redistribution of level populations from $LS$ terms is not necessarily restricted to its corresponding fine-structure levels. A more detailed study is underway.

\citet{BadnellSeaton2003} and OP2005 studied the importance of inner-shell transitions to the opacities, and as previously mentioned in Section~\ref{opsets}, completed the OP opacities with inner-shell photoexcitation and photoionization data determined with the DW method implemented in AS. Among these inner-shell transitions, some arise from K-vacancy levels, and hence, we tested the importance of Auger broadening to the opacities. Present tests confirm the negligible effect previously found by \citet{BadnellSeaton2003} for the 6-element solar mix at the plasma conditions studied.

In the plasma conditions we are considering, collisional broadening dominates and its impact is crucial as seen in the left panel of Fig.~\ref{fig2}. We have tested the effect of broadening on inner-shell processes using the data from OP2005: collisional broadening increases the Rosseland mean by 30\% when compared to only natural (i.e., radiative) and Auger
broadening \citep{Griem1968,Seaton2004}. To further test sensitivity we increased the original collisional broadening by multiplying the collisional widths by factors of 4.5, giving rise to an additional 35\% uplift in the Rosseland mean as found by NP2016, and to 70\% over the original OP2005 if collisional widths are multiplied by a factor 10. Line broadening affects the monochromatic opacity at the most important place; that is, between $u=2.5$ and $u=4.0$ (corresponding to 35 and 55\,Ryd respectively) where the Rosseland weighting function goes through its maximum. The region covered by the Sandia experiment plays a lesser role since the weighting function has dropped significantly there. Relevant questions: are we underestimating collisional broadening? Or, do we have to revise our broadening formulae? Or, are NP2016 overestimating broadening? Furthermore, is there any reason for treating line broadening differently from resonance broadening?

\articlefiguretwo{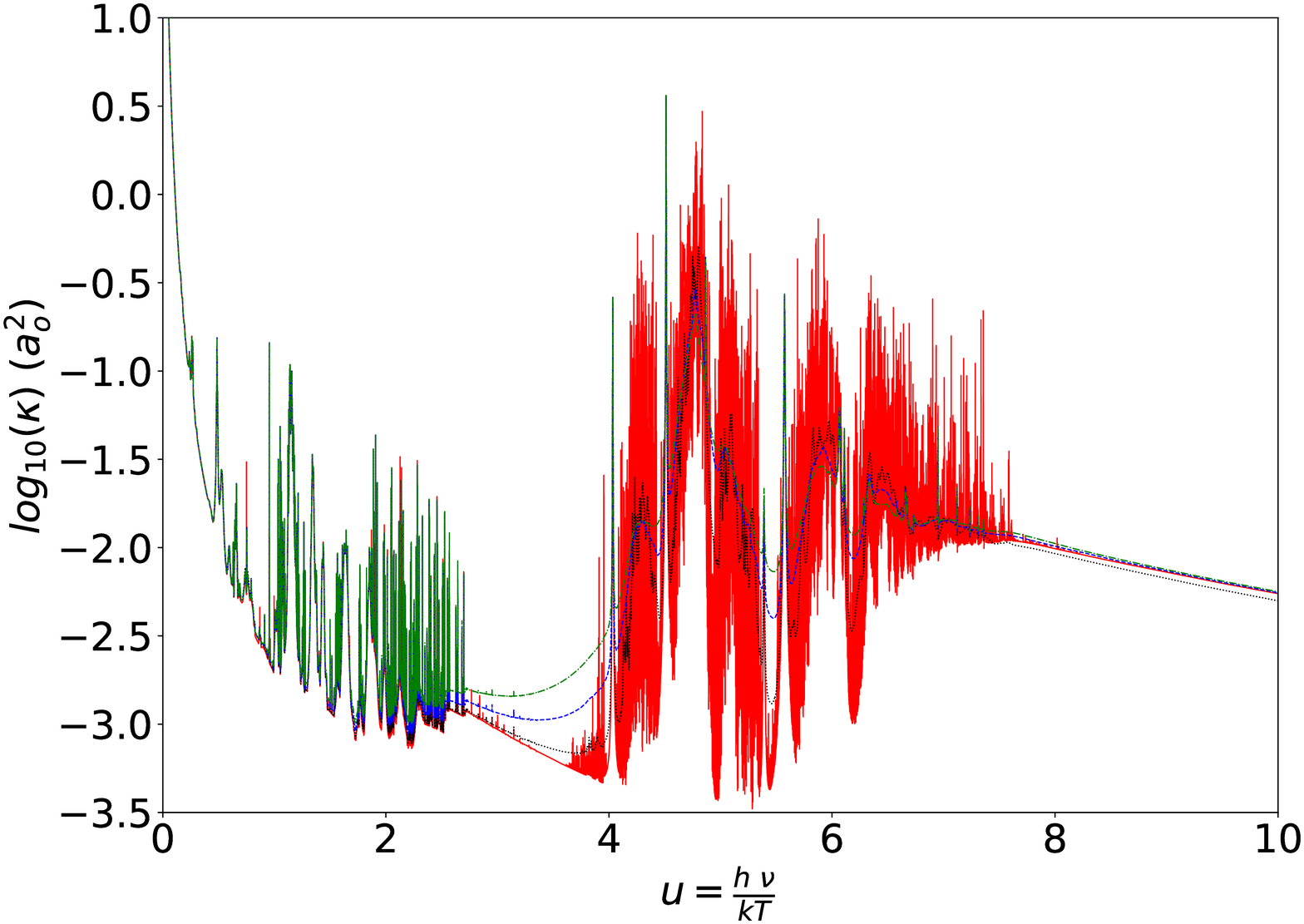}{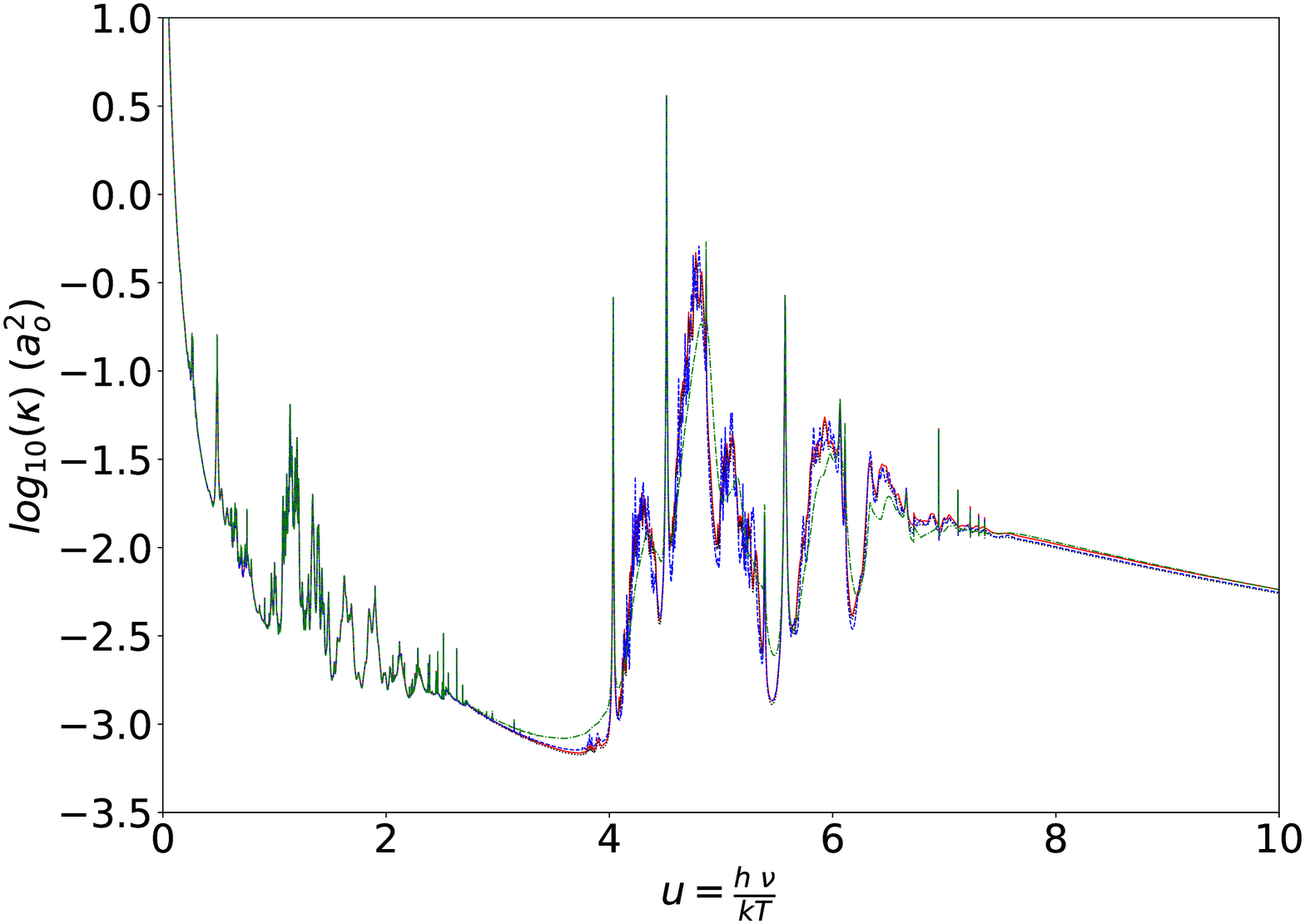}{fig2}{{\it Left}: Effects of collisional broadening. Red curve: no collisional broadening. Black curve: OP2005 with collisional broadening. Blue curve: Line widths multiplied by a factor of 4.5. Green curve: Line widths multiplied by a factor of 10. {\it Right}: Comparison of $\kappa_{PI}$. Blue curve: OP2005 IC. Red curve: OP2005 IC renormalized. Black curve: OP2005 $LS$ renormalized. Green curve: Rerun of inner-shell contributions allowing for CI between complexes (see {\it text} for details).}

The OP2005 inner-shell contribution was estimated ignoring interaction between spectral complexes to make computations more tractable. We have now rerun the same configuration expansion for \ion{Fe}{xvii} as in OP2005 but now allowing for CI between complexes. We find an increase of 10\% in the Rosseland mean opacity for this ion at the plasma conditions of the Sandia experiment.

We continued our study by replacing all the OP2005 valence-shell data (i.e., the OP1995 data) in \ion{Fe}{xvii} by new AS data with an extended set of configurations that allows for photoexcitation to and photoionization from two-hole L-vacancy states. This leads to a further 8\% increase of the Rosseland mean, and if configurations with three-hole L-vacancy states are further taken into account, an overall increase of up to 25\% is found relative to OP2005.

Results are compiled in Table~2. We can see the effects of the three last contributions in Fig.~\ref{fig2}. We should keep in mind that the present results are confined to \ion{Fe}{xvii} in specific physical conditions, namely,  $T=2.1\times10^6$~K and $\rho=3\times10^{22}$~cm$^{-3}$. Such effects may vary for different temperatures, densities, and ionic species.

\begin{table}
  \caption{Effect of extended calculations on the Rosseland mean opacity of \ion{Fe}{xvii} (and \ion{Ni}{xiv} last entry). See {\it text} for details.}\label{tab2}
  \smallskip
  {\centering \small
  \begin{tabular}{ll}
  \tableline
  \noalign{\smallskip}
  Model &  \% Difference\\
  \noalign{\smallskip}
  \tableline
  \noalign{\smallskip}
  OP2005 & --\\
  Resolution $\nu$ mesh -- $10^4$ to $10^5$ points & $+1\%$ \\
   $LS$ to $IC$               & $-2\%$ \\
  Neglecting Auger broadening & $\sim 0\%$\\
  Renormalization & $-2\%$ \\
  Inner shell including CI between $n$ and $n'$ complexes & $+10\%$\\
  Extended set with initial autoionizing states & $+18\%$\\
  + Two- to three-hole inner shell & $+25\%$ \\
  AS {\it vs.} $R$-Matrix for \ion{Ni}{xiv} & $+10\%$ \\
  \noalign{\smallskip}
  \tableline\
  \end{tabular}

  }
\end{table}

\subsection{Conclusions}

We have presented here only preliminary results and extended work is underway, but they are consistent with the TOPAZ and Los Alamos results from Table~1. As for the calculation of raw atomic data, the AS approach is closer to ATOMIC \citep{Colgan2013b, Colgan2013a, Colgan2015, Colgan2016} and some differences are likely to arise from the different sets of configurations/atomic structures used.

In order to understand the discrepancies between the different calculations, a common set of configurations has been selected as a starting point. We are comparing level populations and ionization fractions in an attempt to constrain them to the same values to test the impact of the EOS on the opacities. Then we will go through each step (atomic data, broadening, etc.) to identify and quantify their effect in terms of monochromatic and mean opacities. A new level-resolved $R$-Matrix calculation is underway to analyse the effect of the coupling scheme and to corroborate the accuracy  of the present AS results. A more exhaustive analysis, as well as the study of its impact on stellar models, is under way (Delahaye et al. in preparation).

\acknowledgements FD would like to thank the AFE-Obs. de Paris for support.

\label{References}


\begin{thebibliography}{}
\expandafter\ifx\csname natexlab\endcsname\relax\def\natexlab#1{#1}\fi
\expandafter\ifx\csname url\endcsname\relax
  \def\url#1{\texttt{#1}}\fi
\expandafter\ifx\csname urlprefix\endcsname\relax\def\urlprefix{URL }\fi
\providecommand{\eprint}[2][]{\url{#2}}

\bibitem[{{Asplund} et~al.(2005){Asplund}, {Grevesse}, \&  {Sauval}}]{Asplund2005} {Asplund}, M., {Grevesse}, N., \& {Sauval}, A.~J. 2005, in Cosmic Abundances as Records of Stellar Evolution and Nucleosynthesis, edited by T.~G. {Barnes}, III, \& F.~N. {Bash}, vol. 336 of ASP Conf. Ser., 25

\bibitem[{{Asplund} et~al.(2004){Asplund}, {Grevesse}, {Sauval}, {Allende Prieto}, \& {Kiselman}}]{Asplund2004} {Asplund}, M., {Grevesse}, N., {Sauval}, A.~J., {Allende Prieto}, C., \&   {Kiselman}, D. 2004, \aap, 417, 751

\bibitem[{{Asplund} et~al.(2009){Asplund}, {Grevesse}, {Sauval}, \& {Scott}}]{Asplund2009} {Asplund}, M., {Grevesse}, N., {Sauval}, \&   {Scott} P. 2009, \araa, 417, 751

\bibitem[{Badnell(2011)}]{Badnell2011} Badnell, N.~R. 2011, Comput. Phys. Commun., 182, 1528

\bibitem[{{Badnell} et~al.(2005){Badnell}, {Bautista}, {Butler}, {Delahaye}, {Mendoza}, {Palmeri}, {Zeippen}, \& {Seaton}}]{Badnell2005} {Badnell}, N.~R., {Bautista}, M.~A., {Butler}, K., et al. 2005, \mnras, 360, 458

\bibitem[{Badnell \& Seaton(2003)}]{BadnellSeaton2003} Badnell, N.~R., \& Seaton, M.~J. 2003, J. Phys. B -- At. Mol., 36, 4367

\bibitem[{{Bailey} et~al.(2015){Bailey}, {Nagayama}, {Loisel}, {Rochau}, {Blancard}, {Colgan}, {Cosse}, {Faussurier}, {Fontes}, {Gilleron}, {Golovkin}, {Hansen}, {Iglesias}, {Kilcrease}, {Macfarlane}, {Mancini},   {Nahar}, {Orban}, {Pain}, {Pradhan}, {Sherrill}, \& {Wilson}}]{Bailey2015} {Bailey}, J.~E., {Nagayama}, T., {Loisel}, G.~P., et al. 2015, Nature, 517, 56

\bibitem[{{Berrington} et~al.(1987){Berrington}, {Burke}, {Butler}, {Seaton}, {Storey}, {Taylor}, \& {Yan}}]{ADOC_I}
{Berrington}, K.~A., {Burke}, P.~G., {Butler}, K., et al. 1987, J. Phys. B -- At. Mol., 20, 6379

\bibitem[{Berrington et~al.(1995)Berrington, Eissner, \& Norrington}]{Berrington1995} Berrington, K.~A., Eissner, W.~B., \& Norrington, P.~H. 1995, Comput. Phys.  Commun., 92, 290

\bibitem[{{Blancard} et~al.(2016){Blancard}, {Colgan}, {Coss{\'e}}, {Faussurier}, {Fontes}, {Gilleron}, {Golovkin}, {Hansen}, {Iglesias}, {Kilcrease}, {MacFarlane}, {More}, {Pain}, {Sherrill}, \& {Wilson}}]{Blancard2016} {Blancard}, C., {Colgan}, J., {Coss{\'e}}, P., et al. 2016, Phys. Rev. Lett., 117, 249501

\bibitem[{Blancard et~al.(2012)Blancard, Coss\'e, \& Faussurier}]{Blancard2012} Blancard, C., Coss\'e, P., \& Faussurier, G. 2012, ApJ, 745, 10

\bibitem[{Burke et~al.(1971)Burke, Hibbert, \& Robb}]{Burke1971} Burke, P.~G., Hibbert, A., \& Robb, W.~D. 1971, J. Phys. B -- At. Mol., 4, 153

\bibitem[{{Colgan} et~al.(2013{\natexlab{a}}){Colgan}, {Kilcrease}, {Magee}, {Armstrong}, {Abdallah}, {Sherrill}, {Fontes}, {Zhang}, \& {Hakel}}]{Colgan2013b} {Colgan}, J., {Kilcrease}, D.~P., {Magee}, N.~H., et al. 2013{\natexlab{a}}, High Energ. Dens. Phys., 9, 369

\bibitem[{{Colgan} et~al.(2013{\natexlab{b}}){Colgan}, {Kilcrease}, {Magee}, {Armstrong}, {Abdallah}, {Sherrill}, {Fontes}, {Zhang}, \& {Hakel}}]{Colgan2013a} {Colgan}, J., {Kilcrease}, D.~P., {Magee}, N.~H., Jr., et al. 2013{\natexlab{b}}, in American Institute of Physics Conference Series, edited by J.~D. {Gillaspy}, W.~L. {Wiese}, \& Y.~A. {Podpaly}, vol. 1545 of AIP Conf. Ser., 17

\bibitem[{{Colgan} et~al.(2015){Colgan}, {Kilcrease}, {Magee}, {Abdallah}, {Sherrill}, {Fontes}, {Hakel}, \&  {Zhang}}]{Colgan2015} {Colgan}, J., {Kilcrease}, D.~P., {Magee}, N.~H., et al. 2015, High Energ. Dens. Phys., 14, 33

\bibitem[{{Colgan} et~al.(2016){Colgan}, {Kilcrease}, {Magee}, {Sherrill}, {Abdallah}, {Hakel}, {Fontes}, {Guzik}, \& {Mussack}}]{Colgan2016} {Colgan}, J., {Kilcrease}, D.~P., {Magee}, N.~H., et al. 2016, \apj, 817, 116

\bibitem[{{Cox} \& {Ellers}(1960)}]{Cox1960} {Cox}, A.~N., \& {Ellers}, D.~D. 1960, \aj, 65, 51

\bibitem[{{Cox} \& {Stewart}(1965)}]{Cox1965} {Cox}, A.~N., \& {Stewart}, J.~N. 1965, \apjs, 11, 22

\bibitem[{{Cox} \& {Stewart}(1970{\natexlab{a}})}]{Cox1970a} --- 1970{\natexlab{a}}, \apjs, 19, 243

\bibitem[{{Cox} \& {Stewart}(1970{\natexlab{b}})}]{Cox1970b} --- 1970{\natexlab{b}}, \apjs, 19, 261

\bibitem[{{Cox} \& {Tabor}(1976)}]{Cox1976} {Cox}, A.~N., \& {Tabor}, J.~E. 1976, \apjs, 31, 271

\bibitem[{Cunto et~al.(1993)Cunto, Mendoza, Ochsenbein, \& Zeippen}]{Cunto1993} Cunto, W., Mendoza, C., Ochsenbein, F., \& Zeippen, C.~J. 1993, \aap, 275, L5

\bibitem[{{Delahaye} et~al.(2014){Delahaye}, {Dubernet}, {Zeippen}, \&  {SUPVAMDC Collaboration.}}]{VAMDC2014}
{Delahaye}, F., {Dubernet}, M.~L., {Zeippen}, C.~J., \& {SUPVAMDC Collaboration} 2014, Phys. Scripta, 89, 114005

\bibitem[{{Delahaye} et~al.(2013){Delahaye}, {Palmeri}, {Quinet}, \&   {Zeippen}}]{Delahaye2013}
{Delahaye}, F., {Palmeri}, P., {Quinet}, P., \& {Zeippen}, C.~J. 2013, in EAS Publication Series, edited by G.~{Alecian}, Y.~{Lebreton}, O.~{Richard}, \&   G.~{Vauclair}, vol.~63 of EAS Publ. Ser., 321

\bibitem[{Delahaye \& Pinsonneault(2006)}]{Delahaye2006} Delahaye, F., \& Pinsonneault, M.~H. 2006, ApJ, 649, 529

\bibitem[{{Delahaye} et~al.(2016){Delahaye}, {Zw{\"o}lf}, {Zeippen}, \&   {Mendoza}}]{Delahaye2016} {Delahaye}, F., {Zw{\"o}lf}, C.~M., {Zeippen}, C.~J., \& {Mendoza}, C. 2016, J. Quant. Spectrosc. Ra., 171, 66

\bibitem[{{Dubernet} et~al.(2016){Dubernet}, {Antony}, {Ba}, {Babikov}, {Bartschat}, {Boudon}, {Braams}, {Chung}, {Daniel}, {Delahaye}, {Del Zanna},   {de Urquijo}, {Dimitrijevi{\'c}}, {Domaracka}, {Doronin}, {Drouin}, {Endres},   {Fazliev}, {Gagarin}, {Gordon}, {Gratier}, {Heiter}, {Hill},   {Jevremovi{\'c}}, {Joblin}, {Kasprzak}, {Krishnakumar}, {Leto}, {Loboda},   {Louge}, {Maclot}, {Marinkovi{\'c}}, {Markwick}, {Marquart}, {Mason},   {Mason}, {Mendoza}, {Mihajlov}, {Millar}, {Moreau}, {Mulas}, {Pakhomov},   {Palmeri}, {Pancheshnyi}, {Perevalov}, {Piskunov}, {Postler}, {Quinet},
  {Quintas-S{\'a}nchez}, {Ralchenko}, {Rhee}, {Rixon}, {Rothman}, {Roueff},   {Ryabchikova}, {Sahal-Br{\'e}chot}, {Scheier}, {Schlemmer}, {Schmitt},   {Stempels}, {Tashkun}, {Tennyson}, {Tyuterev}, {Vuj{\v c}i{\'c}}, {Wakelam},   {Walton}, {Zatsarinny}, {Zeippen}, \& {Zw{\"o}lf}}]{VAMDC2016} {Dubernet}, M.~L., {Antony}, B.~K., {Ba}, Y.~A., et al.  2016, J. Phys. B -- At. Mol., 49, 074003

\bibitem[{Eissner et~al.(1974)Eissner, Jones, \& Nussbaumer}]{Eissner1974} Eissner W. , Jones M., \& Nussbaumer, H. 1974, Comput. Phys. Commun., 8, 270

\bibitem[{Fano(1961)}]{Fano1961} Fano, H. 1961, Phys. Rev. A, 124, 1866

\bibitem[{{Griem}(1968)}]{Griem1968} {Griem}, H.~R. 1968, Phys. Rev., 165, 258

\bibitem[{Hibbert(1975)}]{hib75} Hibbert, A. 1975, Comput. Phys. Commun., 9, 141

\bibitem[{Hummer et~al.(1993)Hummer, Berrington, Eissner, Pradhan, Saraph, \& Tully}]{Hummer1993} Hummer, D.~G., Berrington, K.~A., Eissner, W., et al. 1993, A\&A, 279, 298

\bibitem[{{Iglesias} \& {Hansen}(2017)}]{Iglesias2017} {Iglesias}, C.~A., \& {Hansen}, S.~B. 2017, \apj, 835, 284

\bibitem[{{Lynas-Gray} et~al.(1995){Lynas-Gray}, {Seaton}, \& {Storey}}]{LinasGray1995} {Lynas-Gray}, A.~E., {Seaton}, M.~J., \& {Storey}, P.~J. 1995, J. Phys. B -- At. Mol., 28, 2817

\bibitem[{{Mendoza} et~al.(2007){Mendoza}, {Seaton}, {Buerger},   {Bellor{\'{\i}}n}, {Mel{\'e}ndez}, {Gonz{\'a}lez}, {Rodr{\'{\i}}guez},   {Delahaye}, {Palacios}, {Pradhan}, \& {Zeippen}}]{Mendoza2007} {Mendoza}, C., {Seaton}, M.~J., {Buerger}, P., et al. 2007, \mnras, 378, 1031

\bibitem[{{Mondet} et~al.(2015){Mondet}, {Blancard}, {Coss{\'e}}, \&  {Faussurier}}]{Mondet2015} {Mondet}, G., {Blancard}, C., {Coss{\'e}}, P., \& {Faussurier}, G. 2015, \apjs, 220, 2

\bibitem[{{Nahar} \& {Pradhan}(2016{\natexlab{a}})}]{Nahar2016a} {Nahar}, S.~N., \& {Pradhan}, A.~K. 2016{\natexlab{a}}, Phys. Rev. Lett., 116, 235003

\bibitem[{{Nahar} \& {Pradhan}(2016{\natexlab{b}})}]{Nahar2016b} --- 2016{\natexlab{b}}, Phys. Rev. Lett., 117, 249502

\bibitem[{Nussbaumer \& Storey(1978)}]{Nussbaumer1978} Nussbaumer, H., \& Storey, P.~J. 1978, \aap, 64, 139

\bibitem[{{Rogers} \& {Iglesias}(1992)}]{OPAL1992} {Rogers}, F.~J., \& {Iglesias}, C.~A. 1992, \apj, 401, 361

\bibitem[{{Rogers} et al.(1996)}]{Rogers1996} {Rogers}, F.~J., {Swenson} F.~J., \& {Iglesias}, C.~A. 1999, \apj, 456, 902

\bibitem[{{Seaton}(1987)}]{ADOC_II} {Seaton}, M.~J. 1987, J. Phys. B -- At. Mol., 20, 6363

\bibitem[{{Seaton}(1995)}]{Seaton1995} --- 1995, {The Opacity Project} (Bristol, UK: IOP Publishing), vol.~1

\bibitem[{{Seaton} \& {Badnell}(2004)}]{Seaton2004} {Seaton}, M.~J., \& {Badnell}, N.~R. 2004, \mnras, 354, 457

\bibitem[{{Seaton} et~al.(1994){Seaton}, {Yan}, {Mihalas}, \&   {Pradhan}}]{Seaton1994} {Seaton}, M.~J., {Yan}, Y., {Mihalas}, D., \& {Pradhan}, A.~K. 1994, \mnras, 266, 805

\bibitem[{Simon(1982)}]{Simon1982} Simon, N.~R. 1982, \apjl, 260, L87

\bibitem[{Turck-Chi\`eze(1998)}]{TurckChieze1988} Turck-Chi\`eze, S. 1998, Space Sci. Rev., 85, 125

\bibitem[{{Villante} et~al.(2014){Villante}, {Serenelli}, {Delahaye}, \&   {Pinsonneault}}]{Villante2014}
{Villante}, F.~L., {Serenelli}, A.~M., {Delahaye}, F., \& {Pinsonneault}, M.~H. 2014, \apj, 787, 13

\bibitem[{Vinyoles et~al.(2017)Vinyoles, Serenelli, Villante, Basu, Bergström, Gonzalez-Garcia, Maltoni, Peña-Garay, \& Song}]{Serenelli2017} Vinyoles, N., Serenelli, A.~M., Villante, F.~L., et al. 2017, ApJ, 835, 202

\bibitem[{{Zeippen}(1995)}]{Zeippen1995} {Zeippen}, C.~J. 1995, Phys. Scripta, T58, 43

\end{thebibliography}

\end{document}